\documentclass[11pt]{article}
\usepackage{natbib}
\usepackage{url}
\usepackage{amsmath}

\newcommand{\blob}{\circle*{5}}
\newcommand{\oblob}{\circle{5}}

\title{Substitutes for the non-existent square lattice designs for $36$ varieties}
\author{R. A. Bailey, University of St Andrews, UK; \\
Peter J. Cameron, University of St Andrews, UK;\\
L. H. Soicher, Queen Mary University of London, UK;\\
and E. R. Williams, Australian National University}
\date{}

\begin{document}
\maketitle

\begin{abstract}
  Square lattice designs are often used in trials of new varieties of various
  agricultural crops.  However,
  there are no square lattice designs for $36$ 
varieties in blocks of size six for four or more
replicates. Here we use three different approaches to construct designs for up to eight 
replicates.  All the designs perform well in terms of giving a low average variance of
variety contrasts.

Supplementary materials are available online.

\end{abstract}

\section{Introduction}
 In variety-testing programmes, 
later-stage trials can involve multiple replications of up to 100 varieties: 
see \cite{vartrial}. 
Denote the number of varieties by $v$.
Even at a well-run testing centre, variation across the experimental area makes it 
desirable to group the plots (experimental units) into homogeneous blocks, usually too small 
to contain all the varieties. 
As R.~A.~Fisher wrote in a letter in 1938, ``\ldots\ 
on any given field agricultural operations, at least for
centuries, have followed one of two directions'', so that variability among the plots
is well  captured by blocking in one or both of these directions, with no need for more
complicated spatial correlations: see \citet[p.~270]{ben}.
Thus, on land which has been farmed for centuries, 
or where plots cannot be conveniently arranged to allow blocking in two directions 
(rows and columns), it is reasonable to assume the following model
for the yield $Y_\omega$ on plot $\omega$:
\begin{equation}
Y_\omega = \tau_{V(\omega)} + \beta_{B(\omega)} + \varepsilon_\omega.
\label{eq:model}
\end{equation}
 Here $V(\omega)$ denotes the variety planted on $\omega$ and $B(\omega)$ denotes 
the block containing $\omega$.  The variety constants $\tau_i$ are the unknown parameters
 of interest, and the block constants $\beta_j$ are unknown nuisance parameters.
The quantities $\varepsilon_\omega$ are independent identically
distributed random variables with zero mean and common (unknown) variance $\sigma^2$.

For management reasons, it is
often convenient if the blocks can themselves be grouped into replicates,
in such a way that each variety occurs exactly once in each replicate.  Such a
block design is called \textit{resolvable}.
Let $r$ be the number of replicates.

\cite{FY36,FY37} introduced \textit{square lattice designs} for this purpose.
In these, $v=n^2$ for some positive integer $n$, 
and each replicate consists of $n$ blocks of $n$ plots.  
The design is constructed by first listing the varieties in an
abstract $n \times n$ square array $\mathcal{S}$.  The rows of $\mathcal{S}$
form the blocks of the first replicate, and the columns of $\mathcal{S}$
form the blocks of the second replicate.

If $r>2$ then $r-2$ mutually orthogonal Latin squares
$\mathcal{L}_1$, \ldots, $\mathcal{L}_{r-2}$ of order $n$ are needed.
This is not possible unless $r\leq n+1$: see \citet[Chapter 6]{SS}.
For replicate $i$, where $i>2$, superimpose Latin square
$\mathcal{L}_{i-2}$ on the array $\mathcal{S}$:  the $n$ positions where any given
letter of $\mathcal{L}_{i-2}$ occurs
 give the set of varieties in one block.
Orthogonality implies that each block has one variety in common with each block
in each other replicate.
Thus these designs belong to the class of affine resolvable designs defined by
\cite{bose}, and this construction is a special case of that given by \cite{affine}. 
Moreover, all pairwise variety concurrences are in
$\{0,1\}$,
where the \textit{concurrence} of varieties $i$ and $j$ is the number of blocks in
which varieties $i$ and $j$ both occur.  If $r=n+1$ then all pairwise concurrences
are equal to $1$ and so the design is balanced.

Equireplicate incomplete-block designs are typically assessed using the
A-criterion: see \cite{opt}.  
Denote by $\boldsymbol{\Lambda}$ the $v \times v$ \textit{concurrence matrix}: 
its $(i,j)$-entry is equal to the concurrence of varieties $i$ and $j$,
which is $r$ when $i=j$.
The \textit{scaled information matrix}
is $\mathbf{I}-(rk)^{-1}\boldsymbol{\Lambda}$,
where $\mathbf{I}$ is the identity matrix and
$k$ is the block size.  The constant vectors are
in the null space of this matrix.
 The eigenvalues for the other eigenvectors, counting multiplicities,
are the \textit{canonical efficiency factors}. Denote their harmonic mean
as $A$.  
(\cite{JW} call this $E$, but many authors, including several cited in Section~\ref{sec:soma},
use $E$ to denote the smallest canonical efficiency factor.)
Under model~(\ref{eq:model}),
the average variance of the estimator of a difference $\tau_i -\tau_j$
between two distinct varieties is $2\sigma^2/(rA)$.
If the variance in an ideal design with the same number of plots
but no need for blocking is $\sigma_0^2$, then this average variance would be
 $2\sigma_0^2/r$.
Hence $A\leq 1$, and a design maximizing $A$, for given values of $v$, $r$
and $k$, is called \textit{A-optimal}.

\cite{CSCRAB} showed that if $r\leq n+1$ then square lattice designs are
A-optimal (even over non-resolvable designs).

The class of square lattice designs also has some practical advantages.  Adding
or removing a replicate gives another square lattice design, which permits
last-minute changes in the planning stage.  It also means that if a whole
replicate is lost (for example, if heavy rain during harvest flattens the
plants in the last replicate) then the remaining design is A-optimal for its
size.

If $n\in \{2,3,4,5,7,8,9\}$ then there is a complete set of $n-1$ mutually
orthogonal Latin squares of order $n$: see \citet[Chapter~6]{SS}.  
These give square lattice designs
for $n^2$ 
varieties in $rn$ blocks of size $n$ for $r\in\{2, \ldots, n+1\}$.
However, there is not even a pair of mutually orthogonal Latin squares of
order~$6$, so square lattice designs for $36$ 
varieties
are available for
two or three replicates only.
This gap in the catalogue of good resolvable block designs is pointed out in many books:
for example, \cite{CC,JW}.

\cite{PW} used computer search to find an efficient
resolvable design for $36$ 
varieties in
four replicates of blocks of size six.  All pairwise
variety concurrences are in $\{0,1,2\}$.
  It has $A =0.836$, which compares well with the
  unachievable upper bound of $0.840$ for the non-existent square lattice design.

  In this paper we present three new methods of contructing efficient resolvable
  block designs for $36$ 
varieties in $6r$ blocks of size six, for
  $r\in \{4, \ldots, 8\}$.
  These methods are in Sections~\ref{sec:star}--\ref{sec:soma}.  In each case,
  Supplementary Material gives the design as a plain text file, which can easily
  be imported into a spreadsheet or statistical software.

  The \textit{concurrence graph} of an incomplete-block design has a vertex for
  each variety.  The number of edges between vertices $i$ and $j$ is equal to
  the concurrence of varieties $i$ and $j$.
  Although the methods in Sections~\ref{sec:star}--\ref{sec:soma}
  are very different, their designs for eight replicates all have the same
concurrence graph, which we describe in Section~\ref{sec:syl}.
The final sections compare the new designs and discuss further work.
  
\section{The Sylvester graph}
\label{sec:syl}
\label{sec:graph}
The Sylvester graph $\Sigma$ is a graph on $36$ vertices with valency~$5$.
See \cite{DRG} and \cite{RFBweb}.  Here we give enough information about it
to show how the designs in Section~\ref{sec:star} are constructed, using the approach in
\citet[Chapter 6]{CvL}.

Consider the complete graph $K_6$.  It has a set $\mathcal{A}$ of six vertices,
 labelled $1$ to $6$.  There is an edge between every pair of distinct vertices.
Let $\mathcal{B}$ be this set of edges.
A \textit{$1$-factor} is a partition of $\mathcal{A}$ into three edges (subsets of size two).
For example, one $1$-factor consists of the pairs $\{1,2\}$, $\{3,6\}$ and
$\{4,5\}$.  For brevity, we write this  in the slightly non-standard way $12|36|45$
in Table~\ref{tab:D}.  Let $\mathcal{C}$ be the set of $1$-factors.
A \textit{$1$-factorization} is a set of five elements of $\mathcal{C}$ with the property that
 each edge is contained in just one of them.  For example, $d_1$ in Table~\ref{tab:D}
is a $1$-factorization; 
here we use the symbol $||$ to separate the five $1$-factors in~$d_1$.
Let $\mathcal{D}$ be the set of $1$-factorizations.

It was shown by \cite{Syl} that
\begin{itemize}
\item[(a)] $\left | \mathcal{A} \right | = \left | \mathcal{D} \right | = 6$ and
  $\left | \mathcal{B} \right | = \left | \mathcal{C} \right | = 15$;
  \item[(b)] any two elements of $\mathcal{D}$ share exactly one element of $\mathcal{C}$.
  \end{itemize}
(Sylvester used the terms \textit{duads}, \textit{synthemes} and
\textit{synthematic totals} for edges, $1$-factors and $1$-factorizations
respectively.)

\begin{table}
  \caption{The set $\mathcal{D}$ of six $1$-factorizations}
  \label{tab:D}
  \[
\renewcommand{\arraystretch}{1.3}
  \begin{array}{cc}
    d_1: & ||12|36|45||13|24|56||14|35|26||15|23|46||16|25|34||\\
    d_2: & ||12|36|45||13|25|46||14|23|56||15|26|34||16|24|35||\\
    d_3: & ||12|34|56||13|25|46||14|35|26||15|24|36||16|23|45||\\
    d_4: & ||12|34|56||13|26|45||14|25|36||15|23|46||16|24|35||\\
    d_5: & ||12|46|35||13|26|45||14|23|56||15|24|36||16|25|34||\\
    d_6: & ||12|46|35||13|24|56||14|25|36||15|26|34||16|23|45||
  \end{array}
  \]
  \end{table}

Table~\ref{tab:D} shows the six $1$-factorizations, labelled as $d_1$ to $d_6$.
Each of these can be considered as a schedule for a tournament
involving six teams which takes place over five weekends so that each pair of
teams meets exactly once.

\begin{figure}
    \begin{center}
\addtolength{\unitlength}{\unitlength}
\begin{tabular}{c@{\quad}c}
    \begin{picture}(70,70)(-10,0)
      \multiput(0,0)(0,10){7}{\line(1,0){60}}
      \multiput(0,0)(10,0){7}{\line(0,1){60}}
\multiput(5,5)(10,0){6}{\oblob}  
\multiput(5,15)(10,0){6}{\oblob}  
\multiput(5,25)(10,0){6}{\oblob}  
\multiput(5,35)(10,0){6}{\oblob}  
\multiput(5,45)(10,0){6}{\oblob}  
\multiput(5,55)(10,0){6}{\oblob}
\put(5,63){\makebox(0,0){$d_1$}}  
\put(15,63){\makebox(0,0){$d_2$}}  
\put(25,63){\makebox(0,0){$d_3$}}  
\put(35,63){\makebox(0,0){$d_4$}}  
\put(45,63){\makebox(0,0){$d_5$}}  
\put(55,63){\makebox(0,0){$d_6$}}
\put(-3,55){\makebox(0,0){$1$}}  
\put(-3,45){\makebox(0,0){$2$}}
\put(-3,35){\makebox(0,0){$3$}}
\put(-3,25){\makebox(0,0){$4$}}
\put(-3,15){\makebox(0,0){$5$}}
\put(-3,5){\makebox(0,0){$6$}}  
{\multiput(27,53)(0,-20){3}{\line(1,-1){6}}}
{\multiput(27,47)(0,-20){3}{\line(1,1){6}}}
 \end{picture}
&
  \begin{picture}(60,60)
      \multiput(0,0)(0,10){7}{\line(1,0){60}}
      \multiput(0,0)(10,0){7}{\line(0,1){60}}
      {\put(25,35){\blob}
\put(21,35){\makebox(0,0){$a$}}
        \put(15,55){\oblob}
      \put(25,35){\line(-1,2){8.8}}
      \put(55,45){\oblob}
      \put(25,35){\line(3,1){27.8}}
      \put(35,25){\oblob}
      \put(25,35){\line(1,-1){8}}
      \put(5,15){\oblob}
      \put(25,35){\line(-1,-1){18.3}}
      \put(45,5){\oblob}
      \put(25,35){\line(2,-3){18.7}}}
    \end{picture}\\[10pt]
(a) Edges between columns $d_3$ and $d_4$ 
& (b) The starfish centred at vertex $a$
\end{tabular}
\end{center}
    \caption{Some edges  in the Sylvester graph}
  \label{fig:edges}
\end{figure}
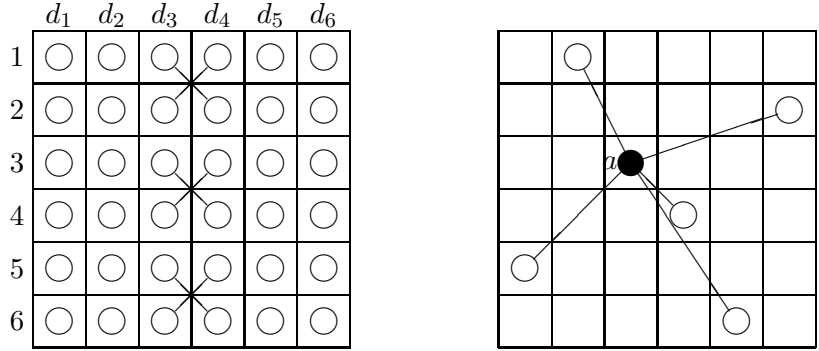

Now we construct the Sylvester graph $\Sigma$ as follows.  The vertex set 
consists of the cells of the $6\times 6$
square array $\mathcal{S}$ with rows labelled by the elements of $\mathcal{A}$ and columns
labelled by the elements of $\mathcal{D}$.  
Given distinct $d_i$ and $d_j$ in $\mathcal{D}$, the
unique $1$-factor they have in common defines six edges in~$\Sigma$, each joining
a vertex in column $d_i$ to a vertex in column $d_j$.  For example, $d_3$ and
$d_4$ have the one-factor $12|34|56$ in common, so we put edges between the
vertices $(1,3)$ and $(2,4)$, $(2,3)$ and $(1,4)$, $(3,3)$ and $(4,4)$, \ldots,
and $(6,3)$ and $(5,4)$, as shown in Figure~\ref{fig:edges}(a).

Thus each vertex in $\Sigma$ is joined to five other vertices, one in each other
row and one in each other column.  Figure~\ref{fig:edges}(b) shows the five edges
at the vertex $a=(3,d_3)$.  We shall call this set of six vertices the
\textit{starfish} centred at $a$.

It can be shown that the graph $\Sigma$ has no triangles or
quadrilaterals.  One consequence of this  is that, given any vertex, 
the vertices at distances one and two from it in the graph are precisely all the other 
vertices in different rows and different columns.  

Denote by $\mathop{\mathbf{Adj}}(\Sigma)$ the adjacency matrix of the graph $\Sigma$.
We call a block design for $36$ varieties in $48$ blocks of size six a \textit{Sylvester design} 
if  there is a permutation of the varieties that takes the concurrence matrix for the design to
$7\mathbf{I} + \mathbf{J} + \mathop{\mathbf{Adj}}(\Sigma)$,
where $\mathbf{J}$ is the all-$1$ matrix.
This means that the concurrences  are $2$ on each edge
of $\Sigma$  and $1$ for every other pair of varieties.
In particular, a Sylvester design is a regular-graph design, 
as defined by \cite{JM}, where the graph is the Sylvester graph $\Sigma$.

Two block designs (for the same set of varieties) 
are \textit{isomorphic} if one can be converted into the
other by a permutation of varieties and a permutation of blocks.
An isomorphism from a block design to itself is called an \textit{automorphism}.

If two block designs are isomorphic then their canonical efficiency
factors are the same and their automorphism groups have the same order,
but neither converse need be true.  In particular, all Sylvester designs have the same
canonical efficiency factors, and hence the same value of the A-criterion, but they are not
all isomorphic.  Of the three that we construct in this paper, no two are isomorphic,
 as we discuss in   Section~\ref{sec:comp}.

\section{New designs constructed from the Sylvester graph}
\label{sec:star}
\subsection{The new designs}
\label{sec:grdes}
Figure~\ref{fig:edges}(a) shows that if we choose two different vertices in
the same column then their starfish will not overlap.  Thus each column gives
what we call a \textit{galaxy} of six starfish, which together include every
vertex just once.  In other words, we can think of a galaxy as a Latin square
of order 6.  Figure~\ref{fig:LS} shows the galaxy of starfish centred on
vertices in column $d_3$.  Just as with a square lattice design, we can
identify the varieties with the $36$ vertices and use
this Latin square to construct a single replicate of six blocks of size six.

\begin{figure}
  \[
  \begin{array}{|c|c|c|c|c|c|}
    \hline
    D & A & B^* & C & E & F\\
    \hline
    F & E & C^* & B & D & A\\
    \hline
    E & B & A^* & D & F & C\\
    \hline
    B & F & D^* & A & C & E\\
    \hline
    A & C & E^* & F & B & D\\
    \hline
    C & D & F^* & E & A & B\\
    \hline
    \end{array}
  \]
  \caption{The galaxy of six starfish defined by column $d_3$: the centre of
    each starfish is marked $*$ and the Latin letters show vertices in the
  same starfish.}
  \label{fig:LS}
  \end{figure}

However, unlike in a square lattice design, the Latin squares defined by different
columns are not orthogonal to each other.  If two vertices are joined by an
edge then they both occur in the two starfish which they define.  Thus if we
use galaxies of starfish from two or more columns then some variety concurrences
will be bigger than one.
On the other hand, a consequence of the lack of triangles and quadrilaterals
is that if two or more galaxies are
used as replicates then there is no other way that two varieties can concur
in more than one block.

We therefore propose the following resolvable designs.
The design $\Gamma_r$ consists of the galaxies of starfish from $r$ columns,
where $0\leq r\leq 6$; $\Gamma_0$ (a design with no blocks) 
and $\Gamma_1$ (a disconnected design) are used in the following
constructions, but are not themselves suitable designs.
For $1\leq r \leq 7$, the design $\Gamma^{\mathrm{R}}_r$ consists of $\Gamma_{r-1}$
together with another replicate whose blocks are the rows of $\mathcal{S}$,
while the design $\Gamma^{\mathrm{C}}_r$ consists of $\Gamma_{r-1}$ together with
another replicate whose blocks are the columns of $\mathcal{S}$.
The design $\Gamma^{\mathrm{C}}_7$ was used by \cite{sesqui}.
For $2\leq r \leq 8$, the design $\Gamma^{\mathrm{RC}}_r$ consists of
$\Gamma^{\mathrm{R}}_{r-1}$
together with another replicate whose blocks are the columns of $\mathcal{S}$.
In particular, $\Gamma^{\mathrm{RC}}_2$ is
the square lattice design whose blocks are the rows and columns.

The automorphisms of $\Sigma$ consist of the symmetric group $S_6$ acting
simultaneously on rows and columns of the array, as well as a further
involution transposing it.  It follows that, for a design consisting of $m$
galaxies (possibly with rows, and possibly with columns), it does not matter
which $m$ galaxies we choose.

When $r=2$ then $\Gamma^{\mathrm{RC}}_2$, $\Gamma^{\mathrm{R}}_2$ and
$\Gamma^{\mathrm{C}}_2$ are square lattice designs, and hence A-optimal, but $\Gamma_2$ is not.
When $r=3$ then $\Gamma^{\mathrm{RC}}_3$ is a square lattice design,
and hence A-optimal,
but none of the others is.  When $r\geq 4$ then
none of the designs is a square lattice design, so we need to calculate the
canonical efficiency factors, and hence $A$.

As discussed in Section~\ref{sec:iso}, for each value of $r$
the designs $\Gamma^{\mathrm{R}}_r$ and $\Gamma^{\mathrm{C}}_r$
have the same canonical efficiency factors, so we do not include
$\Gamma^{\mathrm{R}}_r$ in further comparisons.

Another useful consequence of the lack of triangles and quadrilaterals in $\Sigma$
is that the relations `same row', `same column', `joined in the graph' and `other' form a 
$4$-class association scheme on the set of vertices.  It follows that $\Gamma_6$,
$\Gamma^{\mathrm{C}}_7$ and $\Gamma^{\mathrm{RC}}_8$ are partially balanced with
respect to this association scheme, and so
their canonical efficiency factors can be calculated using the methods in
\cite{AS}.  Table~\ref{tab:PB} shows the results.
In fact,
$\Gamma_6$ and $\Gamma^{\mathrm{RC}}_8$ are also partially balanced with respect
to the $3$-class association scheme obtained by merging the classes `same row' and
`same column'.  Moreover,
$\Gamma^{\mathrm{RC}}_8$ is a Sylvester design.

\begin{table}
  \caption{Canonical efficiency factors and values of the A-criterion for
  the partially balanced designs}
  \label{tab:PB}
  \[
  \renewcommand{\arraystretch}{1.2}
  \begin{array}{ccccccc}
    & & \multicolumn{4}{c}{\mbox{canonical efficiency factors}} & \\
    \multicolumn{2}{r}{\mbox{multiplicity}} & 5 & 5 & 9 & 16 & \\
    r & \mbox{design} & & & & & A\\
\hline
    6 & \Gamma_6 & 1 & 1 & 8/9 & 3/4 & 0.8442\\
    7 & \Gamma^{\mathrm{C}}_7 & 1 & 6/7 & 19/21 & 11/14 & 0.8507\\
    8 & \Gamma^{\mathrm{RC}}_8 & 7/8 & 7/8 & 11/12 & 13/16 & 0.8549
  \end{array}
  \]
  \end{table}

For all the other designs, we calculated $A$ as an exact rational number
by using the \textsf{DESIGN} package \citep{design} in \textsf{GAP} \citep{gap}.
The method used for such exact calculation of block design efficiency measures is 
described in Appendix~B of \citet{SOMA}.
These results were verified by using \textsf{GAP} to find the exact Moore--Penrose
inverse of the 
scaled information matrix, calculate its trace, divide this by $35$, and then invert this
as an exact rational number.

Table~\ref{tab:gap} shows the results to four decimal places.  This shows
that, apart from the square lattice designs  $\Gamma^{\mathrm{RC}}_2$  and
$\Gamma^{\mathrm{C}}_2$, the design
 $\Gamma^{\mathrm{RC}}_r$ always beats $\Gamma^{\mathrm{C}}_r$ and $\Gamma_r$. 
Moreover, $\Gamma^{\mathrm{RC}}_4$ does very slightly better than the design
found by \cite{PW}.

The final column of Table~\ref{tab:gap} shows the value of $A$ for a square
lattice design.  This exists only for $r=2$ and $r=3$, when
$\Gamma^{\mathrm{RC}}_r$ is an example.
For $4 \leq r \leq 7$ there is no square lattice design, so the value shown
gives an unachievable upper bound; in every case, $A$ for
$\Gamma^{\mathrm{RC}}_r$ comes very close to this.

\begin{table}
  \caption{Values of the A-criterion for the designs in Sections~\ref{sec:star}--\ref{sec:soma}} 
  \label{tab:gap}
    \[
  \begin{array}{ccccccccc}
    \hline
    & \multicolumn{3}{c}{\mbox{Section 3}} & \mbox{Section 4} &
\multicolumn{3}{c}{\mbox{Section 5}} &   \mbox{square}\\
    r & \Gamma^{\mathrm{RC}}_r & \Gamma^{\mathrm{C}}_r & \Gamma_r  
    & \Theta_r & \Delta^{\mathrm{RC}}_r & \Delta^{\mathrm{C}}_r & \Delta_r
     &\mbox{lattice}\\
    \hline
    2 &0.7778 &0.7778 & 0.7527 & 0.7778 & 0.7778 &0.7778 & 0.7692 & 0.7778 \\
    3 & {0.8235} & 0.8186 &  0.8091 & 0.8235 & {0.8235} & 0.8219 &  0.8101
    &  {0.8235}\\
    4 & 0.8380 & 0.8341 & 0.8285   & 0.8393&0.8393 & 0.8346 & 0.8292   &  {0.8400}\\
    5 & 0.8453 & 0.8422 & 0.8383 & 0.8464 & 0.8456 & 0.8427 & 0.8383 
    &  {0.8485}\\  
    6 & 0.8498 & 0.8473 & {0.8442} & 0.8510 & 0.8501 & 0.8473 & {0.8442}
     &  {0.8537}\\ 
    7 & 0.8528 & {0.8507} & & 0.8542 & 0.8528 & 0.8507 & &   {0.8571}\\
    8 & {0.8549}  & & & 0.8549 & 0.8549 & &  & \\
    \hline
  \end{array}
  \]
\end{table}

\subsection{Using these new designs}
\label{sec:gruse}
Figure~\ref{fig:whole} shows the design $\Gamma^{\mathrm{RC}}_8$,
starting with the replicates defined by columns and rows.  The varieties are
numbered $1$ to $6$ in row~$1$ of Figure~\ref{fig:edges}(a),
then $7$ to $12$ in row~$2$, and so on. 
For a design with $r$ replicates, use the first two replicates here and any $r-2$
of the others.  A plain-text version of $\Gamma^{\mathrm{RC}}_8$
 is available in the Supplementary Material.

\begin{figure}
\[
\addtolength{\arraycolsep}{-0.7\arraycolsep}
\begin{array}{@{}c@{\quad}c@{\quad}c@{\quad}c}
\mbox{Replicate 1} & \mbox{Replicate 2} & \mbox{Replicate 3} & \mbox{Replicate 4}\\
\begin{array}{cccccc}
1 & 2 & 3 & 4 & 5 & 6\\
7 & 8 & 9 & 10 & 11 & 12\\
13 & 14 & 15 & 16 & 17 & 18\\
19 & 20 & 21 & 22 & 23 & 24\\
25 & 26 & 27 & 28 & 29 & 30\\
31 & 32 & 33 & 34 & 35 & 36
\end{array}
&
\begin{array}{cccccc}
1 & 7 & 13 & 19 & 25 & 31\\
2 & 8 & 14 & 20 & 26 & 32\\
3 & 9 & 15 & 21 & 27 & 33\\
4 & 10 & 16 & 22 & 28 & 34\\
5 & 11 & 17 & 23 & 29 & 35\\
6 & 12 & 18 & 24 & 30 & 36
\end{array}
&
\begin{array}{cccccc}
1 & 2 & 6 & 3 & 4 & 5\\
8 & 7 & 10 & 12 & 11 & 9\\
18 & 16 & 13 & 17 & 15 & 14\\
21 & 24 & 23 & 19 & 20 & 22\\
28 & 29 & 27 & 26 & 25 & 30\\
35 & 33 & 32 & 34 & 36 & 31
\end{array}
&
\begin{array}{cccccc}
2 & 1 & 3 & 5 & 6 & 4\\
7 & 8 & 11 & 10 & 9 & 12\\
15 & 17 & 14 & 18 & 16 & 13\\
23 & 22 & 24 & 20 & 19 & 21\\
30 & 27 & 28 & 25 & 26 & 29\\
34 & 36 & 31 & 33 & 35 & 32
\end{array}
\\
\mbox{}
\\
\mbox{Replicate 5} & \mbox{Replicate 6} & \mbox{Replicate 7} & \mbox{Replicate 8}\\
\begin{array}{cccccc}
3 & 4 & 2 & 1 & 5 & 6\\
10 & 9 & 12 & 11 & 8 & 7\\
14 & 18 & 15 & 16 & 13 & 17\\
19 & 23 & 22 & 21 & 24 & 20\\
29 & 26 & 25 & 30 & 27 & 28\\
36 & 31 & 35 & 32 & 34 & 33
\end{array}
&
\begin{array}{cccccc}
4 & 3 & 5 & 6 & 1 & 2\\
9 & 10 & 7 & 8 & 12 & 11\\
17 & 13 & 16 & 15 & 14 & 18\\
24 & 20 & 21 & 22 & 23 & 19\\
25 & 30 & 26 & 29 & 28 & 27\\
32 & 35 & 36 & 31 & 33 & 34
\end{array}
&
\begin{array}{cccccc}
5 & 6 & 4 & 2 & 3 & 1\\
12 & 11 & 8 & 9 & 7 & 10\\
16 & 14 & 17 & 13 & 18 & 15\\
20 & 21 & 19 & 23 & 22 & 24\\
27 & 25 & 30 & 28 & 29 & 26\\
31 & 34 & 33 & 36 & 32 & 35
\end{array}
&
\begin{array}{cccccc}
6 & 5 & 1 & 4 & 2 & 3\\
11 & 12 & 9 & 7 & 10 & 8\\
13 & 15 & 18 & 14 & 17 & 16\\
22 & 19 & 20 & 24 & 21 & 23\\
26 & 28 & 29 & 27 & 30 & 25\\
33 & 32 & 34 & 35 & 31 & 36
\end{array}
\end{array}
\]
\caption{The design $\Gamma^\mathrm{RC}_8$: columns are blocks}
\label{fig:whole}
\end{figure}

\section{New designs found by computer search}
\label{sec:cyc}

The computer search algorithm used by \cite{PW}
to obtain the efficient design for $36$ 
varieties has been extensively developed, 
both in the range of design types that can be constructed and in the algorithmic approach.  
Significant improvements in computer speed have also facilitated search procedures.  
CycDesigN Version 6.0 \citep{CD}
is a computer package for 
the generation of optimal or near-optimal experimental designs, as measured by the 
A-criterion.  The package has been written in Visual C++ and uses simulated annealing 
in the design search process.  CycDesigN can be used to construct efficient resolvable block 
designs  for $36$ 
varieties in blocks of size six
with a range of values for $r$. Hence running CycDesigN 
separately for $r=3$ through $8$ gives designs $\Theta_r$ with 
the results in Table~\ref{tab:gap}.  
For $r=3$, \ldots, $7$, the design $\Theta_r$ has pairwise 
variety concurrences in 
$\{0,1,2\}$, 
while $\Theta_8$ has concurrences in $\{1,2\}$.  
In fact, $\Theta_8$ is a Sylvester design.
For $r=4$, 
the improvement from 
\cite{PW}, namely $A=0.836$ to that in Table~\ref{tab:gap} ($A=0.839$) is 
representative of overall developments in computer speed and search methods 
throughout the years.

Because $\Theta_r$ is not constructed by simply omitting a replicate from $\Theta_{r+1}$, we
do not show all these designs here. Figure~\ref{fig:emlyn} shows $\Theta_8$.
Plain-text versions for $r=4$, \ldots, $8$ are available in the Supplementary Material.

\begin{figure}
\[
\addtolength{\arraycolsep}{-0.7\arraycolsep}
\begin{array}{@{}c@{\quad}c@{\quad}c@{\quad}c}
\mbox{Replicate 1} & \mbox{Replicate 2} & \mbox{Replicate 3} & \mbox{Replicate 4}\\
\begin{array}{cccccc}
2  & 24 & 13 & 20 & 19 & 22\\         
29 & 9  & 34 & 15 & 36 & 31\\          
18 & 25 & 1  & 16 & 30 & 32\\          
33 & 7  & 14 & 28 & 23 & 5 \\         
6  & 27 & 12 & 3  & 26 & 21\\          
17 & 11 & 8  & 4  & 10 & 35  
\end{array}
&
\begin{array}{cccccc}
32 & 33 & 4  & 5  & 6  & 12\\          
24 & 11 & 29 & 16 & 20 & 2 \\         
30 & 35 & 8  & 36 & 21 & 31\\          
17 & 18 & 27 & 1  & 23 & 19\\          
22 & 26 & 10 & 25 & 13 & 34\\         
28 & 14 & 3  & 7  & 9  & 15
\end{array}
&
\begin{array}{cccccc}
11 & 31 & 5  & 25 & 35 & 13\\          
15 & 10 & 22 & 20 & 8  & 27\\          
4  & 9  & 34 & 24 & 1  & 2 \\         
12 & 26 & 3  & 14 & 29 & 36\\          
21 & 33 & 7  & 19 & 23 & 6 \\         
30 & 16 & 18 & 17 & 28 & 32
\end{array}
&
\begin{array}{cccccc}
30 & 11 & 25 & 31 & 19 & 16\\         
15 & 2  & 10 & 24 & 32 & 34\\         
35 & 20 & 6  & 14 & 18 & 21\\          
13 & 5  & 22 & 4  & 8  & 26\\          
33 & 23 & 1  & 29 & 28 & 27\\          
7  & 3  & 12 & 36 & 9  & 17
\end{array}
\\
\mbox{}
\\
\mbox{Replicate 5} & \mbox{Replicate 6} & \mbox{Replicate 7} & \mbox{Replicate 8}\\
\begin{array}{cccccc}
2  & 8  & 12 & 6  & 35 & 21\\          
4  & 23 & 28 & 30 & 34 & 36\\          
26 & 31 & 5  & 14 & 10 & 15\\          
1  & 7  & 27 & 32 & 13 & 29\\          
9  & 17 & 19 & 16 & 20 & 25\\          
22 & 11 & 33 & 3  & 24 & 18
\end{array}
&
\begin{array}{cccccc}
14 & 20 & 35 & 31 & 15 & 12\\          
5  & 21 & 7  & 26 & 1  & 16\\          
30 & 8  & 6  & 28 & 32 & 2 \\         
29 & 36 & 27 & 25 & 10 & 23\\          
9  & 22 & 19 & 13 & 11 & 24\\          
34 & 33 & 4  & 3  & 17 & 18\\
\end{array}
&
\begin{array}{cccccc}
21 & 27 & 35 & 15 & 23 & 13\\          
28 & 30 & 3  & 24 & 25 & 11\\          
2  & 1  & 36 & 8  & 4  & 16\\          
10 & 20 & 17 & 5  & 32 & 22\\          
14 & 18 & 9  & 6  & 33 & 29\\          
7  & 31 & 12 & 26 & 34 & 19 
\end{array}
&
\begin{array}{cccccc}
29 & 13 & 19 & 2  & 22 & 36\\          
32 & 17 & 33 & 35 & 23 & 11\\          
12 & 5  & 24 & 8  & 27 & 34\\          
20 & 4  & 21 & 25 & 9  & 31\\          
26 & 18 & 3  & 30 & 14 & 28\\          
7  & 10 & 1  & 16 & 15 & 6
\end{array}
\end{array}
\]
\caption{The design $\Theta_8$: columns are blocks}
\label{fig:emlyn}
\end{figure}

\section{New designs constructed from semi-Latin squares}
\label{sec:soma}

Let $\Delta$ be a resolvable block design for $36$ varieties in $6r$ blocks of
size~$6$.  Its \textit{dual} design $\Delta'$ is obtained by interchanging the
roles of blocks and varieties, so it has $6r$ varieties in $36$ blocks of
size~$r$.  If the varieties of $\Delta$ are identified with 
cells of the $6\times 6$ square array $\mathcal{S}$, 
then the blocks of $\Delta'$ also form a $6 \times 6$ square array.  
When each variety in $\Delta'$ occurs exactly once in each
row and once in each column then $\Delta'$ is called a $(6\times 6)/r$
\textit{semi-Latin square}: see \cite{FYSLS} and \cite{PFSLS}.
The term \textit{orthogonal multi-array} is also used: see \cite{brick}.
One way of constructing such a semi-Latin square is to superpose $r$~Latin
squares with disjoint alphabets.  Not all semi-Latin squares arise in this
way, but the resolvability of $\Delta$ forces the $r$~replicates to be a
collection of $r$~Latin squares when $\Delta'$ is a semi-Latin square.

Let $A'$ be the A-criterion for $\Delta'$.  \cite{Roy} proved that
\begin{equation}
  \label{eq:roy}
  \frac{35}{A} = 6(6-r) + \frac{(6r-1)}{A'}.
\end{equation}
Hence $\Delta$ is A-optimal if and only if $\Delta'$ is A-optimal, as \cite{PW2}
showed.

Highly efficient $(6\times 6)/r$ semi-Latin squares have been found by
\cite{brick}, \cite{rabSLS,rabHowell}, \cite{GRSLS} and 
\cite{webSOMA,UniformSLS,SOMA,DGC}.  
In most cases, their duals are not resolvable.  However,
\citet[Section 6]{SOMA} gives an efficient $(6\times 6)/6$ semi-Latin square made by
superposing six Latin squares labelled $L_1$,\ldots, $L_6$.  
These Latin squares can be used to construct
our designs, just as the galaxies in Section~\ref{sec:star}.

For $0\leq r \leq 6$, denote by $\Delta_r$ the design for $36$ varieties in
$6r$ blocks of size six given by the Latin squares $L_1, \ldots, L_r$;
$\Delta_0$ is a design with no blocks, but, for $0 < r\leq 6$,
$\Delta_r$ is the dual of the design called $X_r$ by \citet{SOMA}.
Table~2 of \cite{SOMA} gives $A'$ for $X_2$, \ldots, $X_6$: from this,
$A$ can be calculated from Equation~(\ref{eq:roy}).

As in Section~\ref{sec:star}, we can add to $\Delta_{r-1}$ another replicate
whose blocks are the rows of $\mathcal{S}$, to obtain $\Delta_r^\mathrm{R}$, or
we can add another replicate whose blocks are the columns of $\mathcal{S}$,
to obtain $\Delta_r^\mathrm{C}$.  Adding both of these extra replicates to
$\Delta_{r-2}$ gives $\Delta_r^{\mathrm{RC}}$.  

Unlike the situation in Section~\ref{sec:star}, choosing a different
$r$-subset of $\{L_1,\ldots,L_6\}$ may give a design with a 
value of $A$ different from that for $\Delta_r$,
but computation shows that, in every case, the highest value of the A-criterion
arises from taking $\{L_1,\ldots,L_r\}$ as our $r$-subset. 

Table~\ref{tab:gap} shows the values of the A-criterion for $\Delta_r$,
$\Delta^\mathrm{C}_r$ and $\Delta^\mathrm{RC}_r$, calculated exactly using
the \textsf{DESIGN} package and rounded to four decimal places. We found
that, for each $r=2,\ldots,7$, the canonical efficiency factors of 
$\Delta_r^\mathrm{R}$ are the same as those of 
$\Delta_r^\mathrm{C}$, and so their A-values are the same.  Note that,
just as in Section~\ref{sec:star}, $\Delta_r^{\mathrm{RC}}$ always
beats $\Delta_r^\mathrm{R}$, $\Delta_r^\mathrm{C}$ and $\Delta_r$,
apart from the fact that $\Delta^\mathrm{RC}_2$, $\Delta^\mathrm{R}_2$  and
$\Delta^\mathrm{C}_2$ are all square lattice designs.
Note also that, to four decimal places of the A-criterion, 
$\Delta^\mathrm{RC}_r$ is always at least as good as
$\Gamma^\mathrm{RC}_r$, and sometimes better.

Figure~\ref{fig:wholeDelta} shows the design $\Delta^{\mathrm{RC}}_8$,
with the replicates defined in the order columns, rows, $L_1$, \ldots, $L_6$. 
The varieties are numbered $1$ to $6$ in row~$1$,
then $7$ to $12$ in row~$2$, and so on. 
A plain-text version of this design is available in the Supplementary Material.
For a design with $r$ replicates, use the first $r$ of the given replicates.

\begin{figure}
\[
\addtolength{\arraycolsep}{-0.7\arraycolsep}
\begin{array}{@{}c@{\quad}c@{\quad}c@{\quad}c}
\mbox{Replicate 1} & \mbox{Replicate 2} & \mbox{Replicate 3} & \mbox{Replicate 4}\\
\begin{array}{cccccc}
1 & 2 & 3 & 4 & 5 & 6\\
7 & 8 & 9 & 10 & 11 & 12\\
13 & 14 & 15 & 16 & 17 & 18\\
19 & 20 & 21 & 22 & 23 & 24\\
25 & 26 & 27 & 28 & 29 & 30\\
31 & 32 & 33 & 34 & 35 & 36
\end{array}
&
\begin{array}{cccccc}
1 & 7 & 13 & 19 & 25 & 31\\
2 & 8 & 14 & 20 & 26 & 32\\
3 & 9 & 15 & 21 & 27 & 33\\
4 & 10 & 16 & 22 & 28 & 34\\
5 & 11 & 17 & 23 & 29 & 35\\
6 & 12 & 18 & 24 & 30 & 36
\end{array}
&
\begin{array}{cccccc}
1 & 2 & 3 & 4 & 5 & 6\\
8 & 7 & 12 & 11 & 10 & 9\\
15 & 16 & 13 & 14 & 18 & 17\\
22 & 24 & 23 & 19 & 21 & 20\\
29 & 27 & 26 & 30 & 25 & 28\\
36 & 35 & 34 & 33 & 32 & 31
\end{array}
&
\begin{array}{cccccc}
1 & 2 & 3 & 4 & 5 & 6\\
9 & 11 & 7 & 8 & 12 & 10\\
14 & 13 & 17 & 18 & 16 & 15\\
24 & 21 & 22 & 23 & 20 & 19\\
29 & 28 & 30 & 27 & 25 & 26\\
34 & 36 & 32 & 31 & 33 & 35
\end{array}
\\
\mbox{}
\\
\mbox{Replicate 5} & \mbox{Replicate 6} & \mbox{Replicate 7} & \mbox{Replicate 8}\\
\begin{array}{cccccc}
1 & 2 & 3 & 4 & 5 & 6\\
9 & 10 & 8 & 7 & 12 & 11\\
16 & 13 & 18 & 17 & 14 & 15\\
23 & 24 & 19 & 21 & 22 & 20\\
30 & 29 & 28 & 26 & 27 & 25\\
32 & 33 & 35 & 36 & 31 & 34
\end{array}
&
\begin{array}{cccccc}
1 & 2 & 3 & 4 & 5 & 6\\
10 & 9 & 11 & 12 & 8 & 7\\
17 & 18 & 16 & 15 & 13 & 14\\
20 & 22 & 24 & 19 & 21 & 23\\
27 & 25 & 26 & 29 & 30 & 28\\
36 & 35 & 31 & 32 & 34 & 33
\end{array}
&
\begin{array}{cccccc}
1 & 2 & 3 & 4 & 5 & 6\\
11 & 12 & 10 & 9 & 7 & 8\\
18 & 17 & 14 & 13 & 15 & 16\\
22 & 19 & 23 & 20 & 24 & 21\\
26 & 27 & 25 & 30 & 28 & 29\\
33 & 34 & 36 & 35 & 32 & 31
\end{array}
&
\begin{array}{cccccc}
1 & 2 & 3 & 4 & 5 & 6\\
12 & 10 & 7 & 8 & 9 & 11\\
14 & 15 & 18 & 17 & 16 & 13\\
21 & 23 & 20 & 24 & 19 & 22\\
28 & 30 & 29 & 25 & 26 & 27\\
35 & 31 & 34 & 33 & 36 & 32
\end{array}
\end{array}
\]
\caption{The design $\Delta^\mathrm{RC}_8$: columns are blocks}
\label{fig:wholeDelta}
\end{figure}

Although the design $X_6$
of \citet{SOMA} was not constructed using the Sylvester graph, 
it turns out that $\Delta^\mathrm{RC}_8$ is a Sylvester design, and so
has the same canonical efficiency factors as $\Gamma_8^\mathrm{RC}$ and $\Theta_8$.

\section{Comparison of designs}
\label{sec:comp}

\subsection{Isomorphism}
\label{sec:iso}
We can determine the automorphism
group of a block design and check block design isomorphism using the
\textsf{DESIGN} package \citep{design}. We can also use this package
to check whether two block designs have the same canonical efficiency factors. 
Extended examples of the use of the \textsf{DESIGN} package for the 
construction, classification and analysis of block designs are 
given in \citet{DGC}.

For $r=8$, we have checked that
the designs given in Sections~\ref{sec:star}--\ref{sec:soma}
are all Sylvester designs, so they all have the same canonical efficiency
factors.  However, 
$\Gamma^{\mathrm{RC}}_8$, $\Theta_8$ and $\Delta^{\mathrm{RC}}_8$ have automorphism
groups of order $1440$, $1$ and $144$ respectively,
so no two of these designs are isomorphic.

Now the square lattice design $\Gamma^{\mathrm{R}}_2$ is isomorphic 
to $\Gamma^{\mathrm{C}}_2$,
and we found, using the \textsf{DESIGN} package, that 
$\Gamma^{\mathrm{R}}_7$ is isomorphic to $\Gamma^{\mathrm{C}}_7$.
For $3 \leq r \leq 6$, it turns out that the designs $\Gamma^{\mathrm{R}}_r$ 
and $\Gamma^{\mathrm{C}}_r$ are not isomorphic, but
they have the same canonical efficiency factors.

We found that, for $r=2,3,5,7$, $\Delta^{\mathrm{R}}_r$ is isomorphic to
$\Delta^{\mathrm{C}}_r$. For $r=4,6$, the designs $\Delta^{\mathrm{R}}_r$
and $\Delta^{\mathrm{C}}_r$ are not isomorphic, but 
they have the same canonical efficiency factors.
We do not yet know a theoretical reason for this.

Further, although they are the same up to four decimal places, we found that,
to seven decimal places, the value of the A-criterion for $\Gamma_5$
is 0.8382815, but for $\Delta_5$ this value is 0.8382679. Similarly,
up to seven decimal places, the A-value for $\Gamma_6^\mathrm{C}$
is 0.8472622, but for $\Delta_6^\mathrm{C}$ this value is 0.8472563.
For $\Gamma_7^\mathrm{RC}$, the A-value to seven decimal
places is 0.8527641, but for $\Delta_7^\mathrm{RC}$ this value is
0.8527611.  Designs $\Gamma_6$ and $\Delta_6$ are not isomorphic,
but have the same canonical efficiency factors.  The same holds for
$\Gamma_7^\mathrm{C}$ and $\Delta_7^\mathrm{C}$, and, as we have already
noted, for $\Gamma_8^\mathrm{RC}$ and $\Delta_8^\mathrm{RC}$.

We also found that $\Theta_4$ and $\Delta_4^\mathrm{RC}$ have the same
canonical efficiency factors, but there is no permutation of varieties
taking one concurrence matrix into the other.

\subsection{Robustness}
\label{sec:rob}

If a replicate is lost from one of the designs in 
Section~\ref{sec:star} then
the remaining design is also in 
Table~\ref{tab:gap}.  For example, if the original design
is $\Gamma^{\mathrm{RC}}_5$ then the loss of a replicate leaves
$\Gamma^{\mathrm{RC}}_4$, with $A=0.8380$, in three cases out of five; 
the other two cases leave $\Gamma^{\mathrm{R}}_4$ or $\Gamma^{\mathrm{C}}_4$,
both with $A=0.8341$.  The average
efficiency of the remaining design is $0.8364$, while the worst case is
$0.8341$.

If a replicate is lost from $\Gamma^{\mathrm{RC}}_8$, then $A=0.8528$ in six cases and 
$A=0.8507$ in two cases: the average is $0.8522390$ and the worst case is $0.8507$.  
Losing a replicate from $\Theta_8$ gives eight different values
for $A$, with exactly the same maximum and minimum as those just given; now the average
is $0.8522389$.

If a replicate is lost from one of the designs in
Sections~\ref{sec:cyc}--\ref{sec:soma},
the remaining design need not be in Table~\ref{tab:gap}.  However,
for $4\leq r \leq 8$ we have calculated the worst-case value and the
average value of the A-criterion when a single replicate is lost from
$\Gamma_r^{\mathrm{RC}}$, $\Theta_r$ or $\Delta_r^{\mathrm{RC}}$.
Table~\ref{tab:robust} shows the results, using sufficient decimal places
in each column to show when two values are different.
For $r=4$, 
the worst case for $\Theta_4$ is obtained by deleting the first or fourth replicate,
and the worst case for $\Delta^{\mathrm{RC}}_4$ is obtained by deleting
the first or second replicate:
all four resulting designs are pairwise isomorphic.

\begin{table}
\caption{Worst-case and average values of the A-criterion if a replicate is lost from the 
design shown}
\label{tab:robust}
\renewcommand{\arraystretch}{1.1}
\[
\begin{array}{ccccccc}
\hline
& r & 4 & 5 & 6 & 7 & 8\\
\hline
 & \Gamma_r^{\mathrm{RC}} & 0.8186& 0.8341 &0.8422 & 0.847262& 0.8506638\\
\mbox{worst} & \Theta_r & 0.8219 & 0.8362 & 0.8442 & 0.849411 & 0.8506638\\
& \Delta_r^{\mathrm{RC}} & 0.8219 & 0.8346 & 0.8427 & 0.847256 & 0.8506638\\
\hline
& \Gamma_r^{\mathrm{RC}} &0.8211 & 0.8364  & 0.8443& 0.849047& 0.8522390\\
\mbox{average} & \Theta_r & 0.8227 & 0.8377 & 0.8456 & 0.850595 & 0.8522389\\
& \Delta_r^{\mathrm{RC}} &  0.8227 & 0.8368 & 0.8446 & 0.849040& 0.8522368
\end{array}
\]
\end{table}

\section{Discussion}
\label{sec:discuss}

The authors were very surprised to find that, for $r=8$, the three very
different approaches all produced Sylvester designs.  This leads us to
conjecture that Sylvester designs are A-optimal.  Further evidence for this is 
that, while the first method started from the Sylvester graph, neither of
the others did; indeed, the second method used numerical optimization. 

CycDesigN also calculates upper bounds for the A-criterion.  For $r\leq 7$ these
are same as those shown in the final column of Table~\ref{tab:gap}. For
$r=8$ it gives an upper bound of $0.854931$, compared to the A-criterion for all
the Sylvester designs, which is equal to $0.854929$ to six decimal places.
The proximity of these values gives further support to the conjecture
that the Sylvester designs are A-optimal.

For a given value of $r$, which design should be used in practice?
Table~\ref{tab:gap} shows that the A-values for $\Gamma_r^{\mathrm{RC}}$,
  $\Theta_r$ and $\Delta_r^{\mathrm{RC}}$ are extremely close, but $\Theta_r$
    is always at least as good as the other two.  If the user is concerned
    about the possible loss of one replicate, then Table~\ref{tab:robust}
    shows that $\Theta_r$
    is still at least as good as the other two when $r\leq 7$ but
    $\Gamma_8^{\mathrm{RC}}$ might be preferred to $\Theta_8$.

    If the $36$     varieties are replaced by $36$ treatments consisting of
    all combinations of two factors with six levels
    each, and $r\leq 6$, then levels of these two factors can be identified
    with the blocks of any two replicates which together form a square lattice
    design.  For the designs in Sections~\ref{sec:star} and~\ref{sec:soma},
    these are either rows and columns, or one of rows and columns combined
    with any other replicate.  Table~\ref{tab:gap} shows that the second
    possibility gives higher values for the A-criterion.

\paragraph{Bibliographic note}
An extended abstract for this paper is in \cite{BOO}.

\paragraph{Acknowledgement}
The support from EPSRC grant EP/M022641/1
(CoDiMa: a Collaborative
Computational Project in the area of Computational Discrete Mathematics)
is gratefully acknowledged.


\begin{thebibliography}{9}
\bibitem[Bailey(1990)]{rabSLS}
  Bailey, R.~A. (1990)
  An efficient semi-Latin square for twelve treatments in blocks of size two.
  \textit{Journal of Statistical Planning and Inference} \textbf{26},
  263--266.
\bibitem[Bailey(1997)]{rabHowell}
Bailey, R.~A. (1997)
A Howell design admitting $A_5$.
\textit{Discrete Mathematics} \textbf{167/168}, 65--71.
\bibitem[Bailey(2004)]{AS}
  Bailey, R.~A. (2004)
  \textit{Association Schemes: Designed Experiments, Algebra and
    Combinatorics}. Cambridge University Press, Cambridge.
\bibitem[Bailey and Cameron(2018)]{BOO}
Bailey, R.~A. and Cameron, P.~J. (2018)
Substitutes for the non-existent square lattice designs for $36$ varieties.
\textit{Biuletyn Oceny Odmian} \textbf{35}, 11--13.
\bibitem[Bailey, Cameron and Nilson(2018)]{sesqui}
  Bailey, R.~A., Cameron, P.~J. and Nilson, T. (2018)
  Sesqui-arrays, a generalisation of triple arrays.
  \textit{Australasian Journal of Combinatorics} \textbf{71}, 427--451.
\bibitem[Bailey, Monod and Morgan(1995)]{affine}
Bailey, R.~A., Monod, H. and Morgan, J.~P. (1995)
Construction and optimality of affine-resolvable designs.
\textit{Biometrika} \textbf{82}, 187--200.
\bibitem[Bailey and Royle(1997)]{GRSLS}
  Bailey, R.~A. and Royle, G. (1997)
  Optimal semi-Latin squares with side six and block size two.
  \textit{Proceedings of the Royal Society, Series A} \textbf{453},
  1903--1914.
\bibitem[R.~F.~Bailey(2019)]{RFBweb}
  Bailey, R.~F. (2019)
DistanceRegular.org,
\url{http://www.distanceregular.org/}
(accessed 22 September 2019)
\bibitem[Bose(1942)]{bose}
Bose, R.~C. (1942)
A note on the resolvability of balanced incomplete block designs.
\textit{Sankhy\=a} \textbf{6}, 105--110.
\bibitem[Brickell(1984)]{brick}
  Brickell, E.~F. (1984)
  A few results in message authentication.
  \textit{Congressus Numerantium} \textbf{43}, 141--154.
\bibitem[Brouwer, Cohen and Neumaier(1989)]{DRG}
  Brouwer, A.~E., Cohen, A.~M. and Neumaier, A. (1989)
\textit{Distance-Regular Graphs}.
Ergebnisse der Mathematik und ihrer Grenzgebiete \textbf{318}, 
Springer-Verlag, Berlin-Heidelberg.
\bibitem[Cameron and van  Lint(1991)]{CvL}
  Cameron, P.~J. and van Lint, J.~H. (1991)
  \textit{Designs, Graphs, Codes and their Links}.
  London Mathematical Society Student Texts \textbf{22},
  Cambridge University Press, Cambridge.
\bibitem[Cheng and Bailey(1991)]{CSCRAB}
  Cheng, C.-S. and Bailey, R.~A. (1991)
  Optimality of some two-associate-class partially balanced incomplete-block
  designs.
  \textit{Annals of Statistics} \textbf{19}, 1667--1671.
\bibitem[Cochran and Cox(1957)]{CC}
Cochran, W.~G. and Cox, G.~M. (1957)
\textit{Experimental Designs} (2nd edition).
John Wiley \& Sons, New York.
\bibitem[Fisher et al.(1990)]{ben}
Fisher, R.~A. et al. (1990) 
\textit{Statistical Inference and Analysis: Selected correspondence of R.~A.~Fisher, 
Edited by J.~H.~Bennett}. Clarendon Press, Oxford.
\bibitem[John and Mitchell(1977)]{JM}
John, J.~A. and Mitchell, T.~J. (1977)
Optimal incomplete block designs.
\textit{Journal of the Royal Statistical Society, Series B} \textbf{39}, 39--43.
\bibitem[John and Williams(1995)]{JW}
John, J.~A. and Williams, E.~R. (1995)
\textit{Cyclic and Computer Generated Designs} (2nd edition).
Monographs on Statistics and Applied Probability \textbf{38},
Chapman  \& Hall, London. 
\bibitem[Patterson and Williams(1976a)]{PW}
    Patterson, H.~D. and Williams, E.~R. (1976a)
    A new class of resolvable incomplete block designs.
    \textit{Biometrika} \textbf{63},    83--92.
\bibitem[Patterson and Williams(1976b)]{PW2}
Patterson, H.~D. and Williams, E.~R. (1976b)
Some theoretical results on general block designs.
\textit{Congressus Numerantium} \textbf{15}, 489--496.
\bibitem[Patterson, Williams and Hunter(1978)]{vartrial}
Patterson, H.~D., Williams, E.~R. and Hunter, E.~A. (1978)
Block designs for variety trials.
\textit{Journal of Agricultural Science} \textbf{90}, 395--400. 
\bibitem[Preece and Freeman(1983)]{PFSLS} 
Preece, D.~A. and Freeman, G.~H. (1983)
Semi-Latin squares and related designs.
\textit{Journal of the Royal Statistical Society, Series B}
\textbf{45}, 267--277.
\bibitem[Roy(1958)]{Roy}
Roy, J. (1958)
On the efficiency factor of block designs.
\textit{Sankhy\=a} \textbf{19}, 181--188.
 \bibitem[Shah and Sinha(1989)]{opt}
    Shah, K.~R. and Sinha, B.~K. (1989)
    \textit{Theory of Optimal Designs}.
    Lecture Notes in Statistics \textbf{54},
    Springer-Verlag, New York.
\bibitem[Soicher(2012a)]{webSOMA}
Soicher, L.~H. (2012a)
SOMA Update.
\url{http://www.maths.qmul.ac.uk/~lsoicher/soma} 
(accessed 22 September 2019)
\bibitem[Soicher(2012b)]{UniformSLS}
Soicher, L.~H. (2012b)
Uniform semi-Latin squares and their Schur-optimality. 
\textit{Journal of Combinatorial Designs} \textbf{20}, 
265--277. 
 \bibitem[Soicher(2013a)]{SOMA}
    Soicher, L.~H. (2013a)
    Optimal and efficient semi-Latin squares.
    \textit{Journal of Statistical Planning and Inference} \textbf{143},
    573--582.
\bibitem[Soicher(2013b)]{DGC}
Soicher, L.~H. (2013b)
Designs, groups and computing. 
In \textit{Probabilistic Group Theory, Combinatorics, and Computing. 
Lectures from the Fifth de Br\'{u}n Workshop}, Detinko, A{.}, et al{.} (eds), 
Lecture Notes in Mathematics \textbf{2070}, Springer-Verlag, London, 
pp.~83--107. 
\bibitem[Soicher(2019)]{design}
Soicher,~L.~H. (2019)
The \textsf{DESIGN} package for \textsf{GAP}, Version 1.7.
\url{https://gap-packages.github.io/design} 
\bibitem[Street and Street(1987)]{SS}
Street, A.~P. and Street, D.~J. (1987)
\textit{Combinatorics of Experimental Design}.
Oxford University Press, Oxford.
  \bibitem[Sylvester(1844)]{Syl}
    Sylvester, J.~J. (1844)
    Elementary researches in the analysis of combinatorial aggregation.
    \textit{Philosophical Magazine} \textbf{24}, 285--296.
\bibitem[The \textsf{GAP} Group(2019)]{gap}
The \textsf{GAP}~Group (2019)
\textsf{GAP} -- Groups, Algorithms, and Programming, Version 4.10.2,
\url{http://www.gap-system.org}
\bibitem[VSNI(2016)]{CD}
VSNI (2016)
CycDesigN, Version 6.0, 
\url{http//www.vsni.co.uk/software/cycdesign/}
\bibitem[Yates(1935)]{FYSLS} 
Yates, F. (1935)
Complex experiments.
\textit{Journal of the Royal Statistical Society, Supplement} \textbf{2}, 181--247. 
\bibitem[Yates(1936)]{FY36}
    Yates, F. (1936)
    A new method of arranging variety trials involving a large number of
    varieties.
    \textit{Journal of Agricultural Science} \textbf{226}, 
    424--455.
  \bibitem[Yates(1937)]{FY37}
    Yates, F. (1937)
    A further note on the arrangement of variety trials: quasi-Latin squares.
    \textit{Annals of Eugenics} \textbf{7},
    319--332.
  \end{thebibliography}
\end{document}